%&amstex
\input amstex
\documentstyle{amsppt}
\loadbold
\magnification=\magstep1
{\catcode`\@=11\gdef\logo@{}}
\pagewidth{16 true cm}
\pageheight{24 true cm}
\voffset -5mm
\document
\parindent = 0.5cm
\parskip = 1.5mm
\define\aaa{1.0cm}
\define\bbb{0.8cm}

\define\vek{\bold}
\define\vekgr{\boldsymbol}

\define\ep{\varepsilon}
\define\al{\alpha}

\define\ga{\gamma}
\define\de{\delta}

\define\sig{\sigma}

\define\pa{\partial}
\define\th{\vartheta}
\define\fii{\varphi}
\define\pd#1#2{\frac{\pa#1}{\pa#2}}
\define\po#1{\frac{\pa}{\pa#1}}

\define\la{\lambda}

\define\zee{\zeta}

\pageno=1
\NoRunningHeads
\TagsOnRight

\centerline{\bf LOW ENERGY STRING: AN ARISTOTELIAN TOP?}
\vskip 1cm
\centerline{M. Petr\'a\v s
\footnote{e-mail: petras\@fmph.uniba.sk}}
\vskip .8cm
\centerline{\it Department of Theoretical Physics, Comenius University}
\centerline{\it 842 15 Bratislava, Slovakia}
\vskip 2cm
\indent
    It is argued that a low energy string may be an Aristotelian top,
i.e. a rigid body which however cannot be rotationally excited, but
in an external electromagnetic field exhibits a sort of precession with
a Larmor type angular velocity. On the basis of this observation a
proposal is made for a new two steps derivation of the electroweak
standard model from the string dynamics. In the first step a theory
of Aristotelian top is formulated and studied in more detail and
in the second step an attempt is made to derive the electroweak
standard model from the top dynamics.
Before the symmetry breaking fermions
are represented by straight frozen strings -
rotators, whose symmetry under the
rotation around their axes is interpreted as the group
$U(1)_Y$. The emergence of $SU(2)_L$ group is somewhat less transparent
and is supposed to be connected with the new degree of
freedom of relativistic rotators, which
 leads to the up and down type fermions. The symmetry
breaking is associated with the bending of the rotators under
the influence of Higgs field and with their subsequent transformation
 into the curved frozen strings - tops.
In this new picture of electroweak interaction the chirality of the theory
has a simple and natural explanation, the weak isospin and hypercharge
are inherent properties of relativistic rotators/tops and the
superselection rule associated with electric charge is a consequence of the
accepted distinction between up and down type fermions.
\vskip 3cm
\leftline{PACS: 03.65Pm, 11.25.-w, 12.15.-y, 11.15.Ex}
\newpage
\vskip \aaa
                  \centerline{\bf 1. Introduction}
\vskip \bbb
  String theory is regarded as the most promising
candidate for the unified field theory, including gravity,
free of infinities and capable of solving the problem of phenomenological
constants of the standard model \cite{1}. So far this
program has been realized in a modest extent in the form of
string-derived models such as Calabi-Yau compactification \cite{2}
or orbifold models \cite{3}. One can include here also string-motivated
models as for instance flipped $ SU(5) \times U(1)$ \cite{4}, or string
no-scale supergravity \cite{5}. The main problem on the way to a
"theory of everything" is the selection of true vacuum from
among many possibilities and the determination of the corresponding
expectation values of  generalized Higgs fields.
It is clear that in such a situation it is necessary in modelbuilding
to combine deductive methods with inductive and intuitive ones.

In the present paper a proposal is made for a new two steps derivation of
the electroweak standard model (EWSM) from the string dynamics
(The prospect for derivation of QCD will be shortly discussed in
Conclusions). In the first step (see Fig.1) it is argued that a low
energy string with frozen vibrational and rotational  degrees
of freedom is a top which however cannot be rotationally excited and
consequently has a lowest value of spin, in case of fermions equal
$1/2$.
\vskip .5cm
$$\text{String}$$
$$M_{\text{Pl}} \to \infty \hskip .2cm \swarrow \hskip 1.5cm
  \searrow \hskip .2cm M_{\text{Pl}} \to \infty$$
$$  \text{Aristotelian top} \hskip .3cm \longrightarrow \hskip .3cm
    \text{Standard model}$$
\vskip .3cm
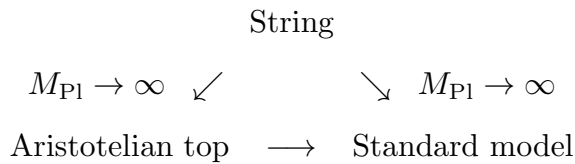
\centerline{Fig.1}
\vskip .5cm
\noindent
 In an external e.g. electromagnetic field  the top exhibits a
sort of precession with Larmor type angular velocity since a precession is
not connected with spin excitation. A top which can rotate in external field
only is an exotic object and will be called the Aristotelian top (AT) in
what follows, in view of the obvious reference to Aristotelian dynamics.

Despite the nonconventional nature of AT its dynamics can be formulated in a
consistent way, as it has been shown in a series of papers
\cite{6}, \cite{7}, \cite{8}, \cite{9}, \cite{10},
where a model of spinning particle realized through AT was elaborated for
the purpose of path integrals. It was shown there that the theory of AT is
much more simple than the theory of conventional top \cite{11}. For instance the
quantization of AT leads to Dirac equation in a straightforward manner, in
contrast to the quantization procedure in case of conventional relativistic
top, where one is confronted with serious problems.

In the second step the final  derivation of EWSM is investigated by a
transition from AT to EWSM under the assumption that nothing substantial has
been lost in the transition. For instance a tacit assumption has been made
that AT has retained those properties of string which are substantial for
the derivation of EWSM. This assumption seems to be satisfied. Actually, as
it is well known, the canonical quantization of strings gives only the
spectrum of various excitations of a free string. To get a more rigorous and
complete quantization one has to apply path integrals, which automatically
lead to a perturbative theory, including hopefully the standard model.

This state of affairs fully corresponds to the situation of quantization of
AT. Here also the canonical quantization leads to the spectrum of free ATs
and does not represent the complete and consistent quantum theory. Due to the
peculiar properties of AT the canonical equations of motion for the rotation
of a free AT are reducible and decouple into separate equations for
canonically conjugated variables $q_i$ and $p_i$ (for the definitions of
canonically conjugated variables for top see Sec. 2). As a result there is
no quantum mechanical correlations between $q_i$ and $p_i$, the principle of
complementarity does not work for free ATs. In order to generate the quantum
dispersions $\vartriangle p_i$, $\vartriangle q_i$ it is necessary to couple
AT to quantized fields and to choose  the values of coupling constants in
such a way as to reach the saturation of the indeterminacy relations $
\vartriangle p_i \vartriangle q_i \geq \hbar$. The free AT becomes then
quantized due to the interaction with the quantum fluctuations of the gauge
fields. In other words a consistent quantization of AT also requires
perturbative quantum field theory in order to generate quantum mechanical
dispersions through the radiation corrections.

In the present paper we restrict ourselves to the fermionic ATs, i.e.
fermionic strings. The gauge as well as Higgs fields will be treated here as
genuine fields necessary for quantization, but their stringy nature will be
ignored.

An important topic which must be discussed here concerns spin of AT/string.
Already the nonrelativistic quantum theory of conventional tops \cite{12} 
revealed
the interesting fact that tops can have both integral and half-integral spins,
provided they are elementary objects, i.e. they have no composite structure.
This condition is essential, since in opposite case the rotation of the top
could be reduced to orbital motion of its constituents (as in the case of
molecules) and the orbital motion cannot lead to half-integral spins. Thus
the condition of elementarity radically restricts candidates among subatomic
objects for a top with half integral spin. In fact the only objects 
which fulfill the condition of elementarity and at the same time
are spatially extended are strings themselves. We thus consider the tops with
spin $1/2$ as fully legitimate representatives of spin $1/2$
particles.

As a result the conventional low energy superstrings do not fall into
the category of ATs considered here, since they involve Grassmannian
variables in their description. Neither the original bosonic strings do,
because they possess only integral angular momenta. We have come to a
surprising conclusion that the upper vertex of the triangle in Fig. 1 seems
to be vacant, there is no string known that would fit the spin $1/2$
AT in the low energy limit. However this conclusion in fact only shows that
our top-down strategy we have followed so far has failed. The reversed
bottom-up strategy dictates that we have to start with AT (if it really
leads to EWSM) and find the corresponding string. In the Appendix B some
arguments will be given that such string does exist. These arguments are
based on a well known theorem from the differential geometry stating that
a curve in 3D space given by parametric equation $\vek x = \vek x (\sigma)$
can be equivalently characterized by a set of equations for three orthogonal
unit vectors $\vek t = \vek t (\sigma)$, $\vek n = \vek n (\sigma)$, 
$\vek b = \vek b (\sigma)$, the tangent, normal and binormal vectors and a
vector $\vek x = \vek x (0)$ which defines the boundary point on the
string. The vectors $\vek t$, $\vek n$, $\vek b$ for any $\sigma$ can be
interpreted as unit vectors defining the body coordinate system of a top. We
can then pass to Euler angles determining the configuration of the top in
space: $\theta = \theta (\sigma)$, $\varphi = \varphi(\sigma)$, $\psi =
\psi(\sigma)$. Thus we can choose either original kinematics for the
description of the string, $\vek x = \vek x (\sigma)$, in which the string
is sequence of points, or a new kinematics, in which a string is a sequence
of tops. Clearly one can expect that in the second case the string would have
both integral and half-integral angular momenta. In the present paper we will
not try to derive the string equations of motion in the new kinematics since
the more urgent task is to ascertain whether QCD follows from AT, too. Also
the issue of the coupling constants calculation can be treated in the
framework of AT without the explicit reference to strings.

The paper consists of three parts. In the first, introductory, part
(Sec. 2 and 3) a nonrelativistic theory of AT is considered. It shows how to
introduce canonically conjugate variables for a top (since Lagrangian theory
of AT does not exist) and also it shows the origin of $U(1)_Y$ symmetry. The
second part (Sec. 4, 5, 6) is devoted to the relativistic theory of free
massless and massive tops. It is shown that a very suitable framework for
this theory represents the dynamical group $SO(3,3)$. Namely this group
contains as a subgroup the group of right-handed and left-handed rotations
of the top
$SO(3)\times SO(3)$, as well as the Lorentz group $SO(3,1)$. Note that
since the top involves for its description Euler angles the Lorentz
transformations of Euler angles follow from the mathematical formalism
developed here. The third part (Sec. 7, 8, 9,10) contains an attempt to derive
EWSM from AT. First, the relativistic version of Sec. 3 is given. Then it is
shown that there are two kinds of rotators/tops, which differ in the Lorentz
transformation properties of Euler angles. It is assumed that they
correspond to up and down type fermions. The assumption is based on the
existence of a superselection rule associated with these two sorts of
rotators/tops. In Sec. 10 the origin of the group $SU(2)_L\times U(1)_Y$ is
discussed from the point of view of Dirac equation. Finally it is argued that
the simplest mechanism for breaking the gauge symmetry in our approach is
the Aristotelian deformation of rotator (i.e. bending without vibration).
However no definite model for the description of this process can be offered
so far.    
\vskip \aaa
        \centerline{\bf 2. Rotators with integral spins}
\vskip \bbb
 In this and the next Section a nonrelativistic model of AT will be developed in
order to show its relationship to EWSM, in particular to the generation of
gauge symmetries, which will be described here by $U(1)_A\times U(1)_Z$
group. The most important property of this model is that it points to the
possible origin of $U(1)_Z$ (that is $U(1)_Y$) symmetry, which will be
associated with the symmetry of straight string - rotator - under the
rotations around its axis. This conjecture is supported by two arguments.
First, it will be shown that this rotational symmetry has a gauge nature
and, second, the corresponding interaction with the gauge field
violates parity.  

          The main problem in the interpretation of fermions
as rotators before symmetry breaking is the integral spin of
the latter. Clearly it is necessary to modify the dynamics of rotators          
 in such a way that half-integral spins are not forbidden.
This necessity is underlined by the fact that the
rotator is a particular case of a top. How is it
possible that tops can possess both integral and half-integral
spins, while rotators only integral ones ? To answer this question
we start with Lagrangian of the free nonrelativistic
symmetric top
$$L=\frac{I_1}{2}(\dot\fii^2\sin^2\theta+\dot\theta^2)+\frac{I_3}{2}
(\dot\fii\cos\theta+\dot\psi)^2\tag2.1$$
where $I_1,I_3$ are moments of inertia. Putting $I_3=0$ we obtain
Lagrangian for the rotator
$$L=\frac{I}{2}(\dot\fii^2\sin^2\theta+\dot\theta^2)\tag2.2$$
with $I_1=I$. This implies the primary constraint 
$$p_\psi=\frac{\partial L}{\partial\dot\psi}=0\tag2.3$$
The canonical Hamiltonian is
$$H_C=p_\theta\dot\theta+p_\fii\dot\fii+p_\psi\dot\psi-L=$$
$$=\frac{1}{2I}(p_\theta^2+\frac{1}{\sin\theta}p_\fii^2)\tag2.4$$

          There are two alternatives how to proceed further:
either one suppresses the variables $\psi,p_\psi$  completely and
restricts to angles $\theta,\fii$ only, or one retains formally
the original 3-dimensional configuration space, in which 
however $\psi$  plays now the role of a gauge variable. We shall
follow the second route, since we want to maintain the contact
with the top dynamics as close as possible. In that case
the constraint $\thetag{2.3}$ is taken in the weak sense, i.e. as an
initial condition.

         Instead of canonical Hamiltonian $H_C$ one introduces
in this case the extended Hamiltonian
$$H=H_C+a(t)p_\psi\tag2.5$$
where $a(t)$ is an arbitrary function, reflecting the fact that
we have to do with a gauge system. The generating function of
an infinitesimal gauge transformation is
$$F=p_\psi\de\al(t)$$
The transformed Hamiltonian reads
$$H'=H+\frac{\partial F}{\partial t}=H_C+a'(t)p_\psi$$
where $a'(t)=a(t)+\de\dot\al(t)$. Note also that
$$\psi'=\psi+\{F,\psi\}=\psi+\de\al(t)$$

 In quantum theory the constraint $\thetag{2.3}$ implies
$$\frac{\partial}{\partial\psi}\Phi(\theta,\fii,\psi)=0\tag2.6$$
When taken as an initial condition $\thetag{2.6}$ holds in fact for any
$t$, because the operator $\frac{\partial}{\partial\psi}$ commutes with 
the operator form
of the Hamiltonian $\thetag{2.5}$. The wave function $\Phi$  of the top
can be expressed by means of the functions $D^s_{mk}(\theta,\fii,\psi)$,
the matrix elements of an irreducible representation of the
SO(3) group. The constraint $\thetag{2.6}$ can be fulfilled for $k=0$ only,
which implies an integral spin $s$. We conclude that the integral
spin of rotators is a consequence of $\thetag{2.3}$, which follows
from the particular form of Lagrangian $\thetag{2.2}$.

         The classical model of spinning particle elaborated
in \cite{6} involves non-standard approach in that this model cannot
be described by a Lagrangian. Due to the absence of the rotational
kinetic energy in the Hamiltonian the spinning particle
is a non-Lagrangian object, but can be described in the framework
of the standard canonical formalism. The canonical Hamiltonian,
including orbital degrees of freedom, is
$$H_C=\frac{1}{2m}(\vek p-\frac{e}{c}\vek A)^2+eA_0-\vekgr\mu.\vek B\tag2.7$$
where $\vekgr\mu=\frac{e}{mc}\vek s$ is the magnetic moment
and the components of spin $\vek s$ are
$$s_1=\xi_2\eta_3-\xi_3\eta_2+\xi\frac{\xi_1}{\xi^2_1+\xi^2_2}\xi_i\eta_i$$
$${s_2=\xi_3\eta_1-\xi_1\eta_3+\xi\frac{\xi_2}{\xi^2_1+\xi^2_2}\xi_i\eta_i}$$
$${s_3=\xi_1\eta_2-\xi_2\eta_1}\tag2.8$$
where $\xi_i,\eta_i$ are canonically conjugated variables and $\xi_i$  are
defined by means of Euler angles
$$\xi_1=e^\psi\sin\theta\sin\fii$$
$$\xi_2=-e^\psi\sin\theta\cos\fii$$
$$\xi_3=e^\psi\cos\theta$$
$$\xi=\sqrt{\xi_i^2}= e^{\psi}\tag2.9$$
Since the Hamiltonian $\thetag{2.7}$ does not involve rotational kinetic
energy it does not contain moments of inertia. Therefore
conventional transition to rotator ($I_3=0$) is no longer possible.
      
  Despite this fact one can realize the rotator mode
of the top by introducing the gauge symmetry as in $\thetag{2.5}$
$$H=H_C+a(t)s_3'=$$
$$= \frac{1}{2m}(\vek p-\frac{e}{c}\vek A)^2+eA_0-\vekgr\mu.\vek B+a(t)s_3'\tag2
.10$$
where $s_3'=\nu_{3i}s_i=\xi_i\eta_i$ is the projection of spin to the
symmetry axis parallel to the unit vector $\nu_{3i}=\frac{\xi_i}{\xi}$. The
generating function for the gauge symmetry is again $F=\de\al(t)s_3'$
and the new Hamiltonian
$$H'=H+\{H,F\}+\frac{\partial F}{\partial t}=H+\de\dot\al(t)s_3'$$
has the same form as the old one due to the vanishing Poisson
brackets
$$\{s_i,s_3'\}=0\tag2.11$$
The arbitrary function $a(t)$, which transforms according to equation
$$a'(t)=a(t)+\de\dot\al(t)\tag2.12$$
has formally the status of a physical variable. As a result the
action S must be varied with respect to this variable and one
obtains
$$\frac{\partial H}{\partial a}=s_3'=0\tag2.13$$
i.e. the same constraint as in the case of the conventional
rotator and with the same conclusion concerning the integral spins.

     In the next section we shall show that in order to
get half-integral spin it is necessary, in addition to the choice
of non-conventional Hamiltonian $\thetag{2.10}$, to change also the 
interpretation of $a(t)$ and $\al(t)$.

\vskip\aaa
\centerline{\bf 3. Rotators with integral and half-integral spins}
\vskip\bbb

In order to reformulate the gauge transformations
for the nonrelativistic rotator it is useful to start with
the corresponding action integral
$$S=\int^{t_2}_{t_1}(\eta_i\dot\xi_i-R)dt-\xi_{2i}\eta_{2i}\tag3.1$$
where $R$ is the Routh function \cite{6}
$$R=-\frac{1}{2}m\vek v^2+eA_0-e\frac{\vek v}{c}.\vek A-
\vekgr\mu.\vek B+as_3'\tag3.2$$
This function plays the role of a Hamiltonian with respect to
rotational coordinates $\xi_i,\eta_i$  and the role of a Lagrangian
(with the opposite sign) with respect to orbital coordinates.
As mentioned before, our top is a non-Lagrangian object
and cannot be characterized by a pure Lagrangian.

       In the new approach to gauge transformations we shall
assume that the particle interacts with some vector field $Z^\mu$
in addition to the electromagnetic field $A^\mu$ and that
$$a(t)=-\frac{g}{\hbar}(Z^0-\frac{\vek v}{c}.\vek Z)\tag3.3$$
where $Z^{\mu}$ is taken in the point $\vek x=\vek x(t)$, the position of the
particle. Furthermore we put
$$\al(t)=-\varLambda(\vek x(t),t)$$
where $\varLambda(x)$ is an arbitrary function of x. The gauge transformation
\thetag{2.12} leads now to the conventional vector potential transformation
$$Z^{\mu '}=Z^\mu+\frac{\hbar c}{g}\frac{\partial \varLambda}{\partial x_\mu }
                                                                    \tag 3.4$$
Note that the original meaning of $a(t)$ and $\al(t)$ as
arbitrary functions of $t$ remains preserved: arbitrariness of $a(t)$
is for instance associated with the arbitrariness proper
to any gauge field characterized by a vector potential.

The decisive moment of the new interpretation of gauge
transformation is connected with the introduction of a new term
in the action integral, namely the term
$$S_{field}=-\frac{1}{4}\int G_{\mu\nu}G^{\mu\nu}dVdt\tag3.5$$
where
$$G_{\mu\nu}=\partial_\mu Z_\nu-\partial_\nu Z_\mu$$
This term is an obvious consequence of the new interpretation
of $a(t)$ as given by \thetag{3.3}. The formal status of $a(t)$ as a physical
variable is now transferred to $Z^\mu(x)$. Variation of $S$ with
respect to this function would lead without \thetag{3.5} to inconsistencies.
After inclusion of $S_{field}$ this variation leads to the
field equations for $Z^{\mu}$. The Coulomb law following from these
equations
$$\int_S\vek G.d\vek s=\frac{g}{\hbar}s_3'\tag3.6$$
is now the correct replacement of \thetag{2.13} as a gauge constraint.
The new constraint \thetag{3.6}, unlike \thetag{2.13}, can be fulfilled in
quantum theory both for integral and half-integral spins, since
it is a condition on the field $G$ and not on $s_3'$.

So we come to the conclusion that half-integral spin
of rotators is possible under two conditions: i. rotators do not
have rotational kinetic energy, ii. the function $a(t)$ in \thetag{3.2}
must be interpreted as \thetag{3.3}, i.e. an interaction with a new
vector field must be introduced.

Note that a finite gauge transformation reads
$$\xi_i'=e^{-\varLambda}\xi_i$$
$$\eta_i'=e^\varLambda\eta_i\tag3.7$$
The action integral \thetag{3.1} with
$$R=-\frac{1}{2}m\vek v^2+eA_0-e\frac{\vek v}{c}.\vek A-
\vekgr\mu.\vek B-\frac{g}{\hbar}s_3'(Z^0-\frac{\vek v}{c}.\vek Z)\tag3.8$$
where $g/\hbar$ is a coupling constant, is invariant with respect to
\thetag{3.7}, since
$$\eta_i\dot\xi_i\to\eta_i\dot\xi_i-s_3'(\frac{\partial\varLambda}
{\partial t}+\frac{\partial\varLambda}{\partial\vek x}.\vek v)\tag3.9$$
and the last term on the right hand side of \thetag{3.9} cancels with the term
coming from $R$. Note that the last term in \thetag{3.8} violates parity,
since $ s_3^, $ is a pseudoscalar - a projection of spin $ \vek s $ on the
symmetry axis, $ s_3' = \vekgr\nu_3 . \vek s $.

Summarizing the results of the last two Sections we can say that the
fermionic nature of rotators requires that the rotational symmetry be
realized through the gauge symmetry $ U(1)_{Z} $, assuming that rotator
interacts with some vector field $Z^{\mu} $. Taking into account the
standard interaction with electromagnetic field $ A $ we see that the total
gauge symmetry is $ U(1)_A \times U(1)_Z $.

\vskip \aaa
\centerline{\bf 4. Classical theory of massless tops}
\vskip \bbb

In \cite{6} the classical theory of point-like massive
tops was elaborated. Here we shall show that this theory can
be easily extended to massless tops. The starting point is
the group $SO(3,3)$, which reflects the fact that the most general
motion of a top can be regarded as right-handed and left-handed
rotations. The generators of this group $S^{AB},A,B=0,1,2,3,4,5$
obey the Poisson brackets relations
$$\{S^{AB},S^{CD}\}=g^{BD}S^{AC}+g^{AC}S^{BD}-
g^{BC}S^{AD}-g^{AD}S^{BC}\tag4.1$$
where $diag\ g^{AB}=1,-1,-1,-1,1,1$. The expressions for $S^{AB}$
are given in Appendix A.
       The standard approach to spinning particles is based
on the representation of Poincar\'e group. On the classical level
the generators of this group are $p^\mu$ and $M^{\mu\nu}=
x^\mu p^\nu-x^\nu p^\mu+S^{\mu\nu}$.
The Casimir invariants are $p^2$ and $w^2$, where $w^{\mu}$ is Pauli-Lubanski
vector
$$w_\sig=\frac{1}{2}\ep_{\la\mu\nu\sig}p^\la S^{\mu\nu}\tag4.2$$

In our case the spinning particle is realized through top, so
that besides $p^2$ and  $w^2$ other invariants exist, namely
$$U=S^{45},\qquad V=p_\mu S^{\mu 4},\qquad W=p_{\mu} S^{\mu 5}\tag4.3$$
for which it holds
$$\{M^{\mu\nu},U\}=0,\qquad \{M^{\mu\nu},V\}=0,\qquad
\{M^{\mu\nu},W\}=0\tag4.4$$
One can also show that
$$\{w^\mu,U\}=0,\qquad \{w^\mu,V\}=0,\qquad \{w^\mu,W\}=0\tag4.5$$
When $p^2=m^2\neq 0$, the quantities $U,V/m,W/m$ form the Lie algebra
of left-handed rotations of the top. In case $p^2=0$  one can derive
the relations
$$\{V,W\}=0,\qquad \{V,U\}=-W,\qquad \{W,U\}=V\tag4.6$$
From \thetag{4.6} it follows that the group of left-handed motions of
the massless top is $E(2)$, the group of motions in the Euclidean
plane. The group of right-handed motions is given by the little
Lorentz group, which is conventionally defined for standard
momentum $p=p(1,0,0,1)$. From the definition of $w^\mu$ we have
$$w^0=\vek p.\vek s,\qquad \vek w=p^0\vek s-\vek p\times\vek N$$
where
$$\vek s=(S^{23},S^{31},S^{12})$$
$$\vek N=(S^{01},S^{02},S^{03})$$
and one obtains
$$w^0=ps_3$$
$$w_1=p(s_1+N_2)$$
$$w_2=p(s_2-N_1)$$
$$w_3=ps_3\tag4.7$$
The Lie algebra of right-handed motions corresponds also to $E(2)$
$$\{w_1,w_2\}=0,\qquad \{w_1,s_3\}=w_2,\qquad \{w_2,s_3\}=-w_1\tag4.8$$
Casimir invariants of both algebras are $V^2+W^2$ and $w_1^2+w_2^2$,
respectively. We shall show that
$$V^2+W^2=w_1^2+w_2^2\tag4.9$$
For this purpose we first introduce complex vector
$$U^\mu=-S^{\mu 4}+iS^{\mu 5}$$
From the definitions of $S^{AB}$ given in Appendix A one can derive
for $U^\mu$ and complex conjugate vector $U^{\mu*}$ the following relations
$$U^{\mu*}U^\nu+U^{\nu*}U^\mu=2S^{\mu\sig}S_\rho^{\ \nu}\tag4.10$$
Denoting $P=p_\mu U^\mu$ one gets from \thetag{4.10}
$$V^2+W^2=PP^*=p_\mu p_\nu S^{\mu\rho}S_\rho^{\ \nu}\tag4.11$$
From the definition of $w^\mu$ it follows
$$-w^2=\frac{1}{2}S^{\mu\nu}S_{\mu\nu}p^2 + p_\mu p_\nu S^{\mu\rho}S_\rho^{\ \nu
}
\tag4.12$$
Using the expressions for $S^{\mu\nu}$ given in Appendix A one becomes
$$S^{\mu\nu}S_{\mu\nu}=2U^2\tag4.13$$
If $p^2=m^2\neq 0$, then eq. \thetag{4.11} - \thetag{4.13} imply
$$-\frac{w^2}{m^2}=U^2+\frac{V^2}{m^2}+\frac{W^2}{m^2}\tag4.14$$
The last relation reflects the fact that the laboratory and
the body components of spin lead to the same absolute value of
spin.

       In case of massless particle one gets instead of
\thetag{4.14} the equation
$$-w^2=V^2+W^2\tag4.15$$
If one takes into account that for $p^\mu=p(1,0,0,1)$ one has
$w^0=w_3$, then from \thetag{4.15} one obtains \thetag{4.9}.

\vskip \aaa
\centerline{\bf 5. Quantum theory of massless tops}
\vskip \bbb

       We summarize first the quantum theory of massive tops
given in \cite{7}. Quantum mechanical generators of $SO(3,3)$ group
are defined in Appendix A. They can be obtained from their
classical expressions by the conventional replacement $\eta_i\to\frac{\partial}
{\partial\xi_i}$ up to the ordering ambiguity (for simplicity the factor
$-i$ is
omitted) and they obey the commutation relations
$$[S^{AB}S^{CD}]=g^{BD}S^{AC}+g^{AC}S^{BD}-g^{BC}S^{AD}-g^{AD}S^{BC}\tag5.1$$
The ordering problem was solved in \cite{7} by means of the requirement
that the representation of $SO(3,3)$ be finite dimensional (in order to
get the conventional finite dimensional representations of Lorentz group).
This leads to tops
with some spin $s$ (integral or half-integral). Taking $s=1/2$
one gets tops, which can assume unique spin value characteristic
for conventional fermionic particles (leptons and quarks).

The generators of Poincar\'e group are
$$p^\mu=i\partial^\mu,\qquad M^{\mu\nu}=i(x^\mu\partial^\nu-x^\nu\partial^\mu)
-iS^{\mu\nu}$$
Components of Pauli-Lubanski vector of a massless particle for
standard momentum $p^\mu=p(1,0,0,1)$ have the form similar to \thetag{4.7}
$$w^0=pI_3$$
$$w_1=p(I_1+N_2)$$
$$w_2=p(I_2-N_1)$$
$$w_3=pI_3\tag5.2$$
where $I_i$ can be obtained from the classical expressions \thetag{2.8}
by means of the replacement $\eta_i\to\frac{\partial}{\partial\xi_i}$ and
$\vek N$ is
defined as follows
$$\vek N=-\vekgr\nu_3\times\vek I+s\vekgr\nu_3\tag5.3$$
Here the term $s\vekgr\nu_3$ comes from the ordering ambiguity and we assume
that $s=1/2$.

As in the classical case Lie algebra of  $E(2)$ group can be derived
$$[w_1,w_2]=0,\qquad [w_1,I_3]=w_2,\qquad [w_2,I_3]=-w_2\tag5.4$$
The quantum analogues of \thetag{4.3} are
$$U=\vekgr\nu_3.\vek I,\qquad V=-\frac{1}{2}(P+P_c),\qquad
W=\frac{1}{2i}(P-P_c) \tag5.5$$
where $P=p_\mu U^\mu,\quad P_c = p_\mu U^{\mu}_c$ and
$$U^0=\vekgr\nu.\vek I$$
$$U^{0}_c=\vekgr\nu^*.\vek I$$
$$\vek U=-i\vekgr\nu\times\vek I+is\vekgr\nu$$
$$\vek U_c=i\vekgr\nu^*\times\vek I-is\vekgr\nu^*\tag5.6$$
with $\vekgr \nu $ defined in Appendix A.
In analogy with \thetag{4.4} and \thetag{4.5} it holds
$$[M^{\mu\nu},U]=0,\qquad [M^{\mu\nu},V]=0,\qquad [M^{\mu\nu},W]=0\tag5.7$$
$$[w^\mu,U]=0,\qquad [w^\mu,V]=0,\qquad [w^\mu,W]=0\tag5.8$$
The left-handed motions of the massless top are given also by
the $E(2)$ group
$$[V,W]=0,\qquad [V,U]=-W,\qquad [W,U]=V\tag5.9$$
The quantum analogues of eq. \thetag{4.10} can be derived from the definitions
\thetag{5.6} and the results reads
$$U^{\mu}_cU^\nu+U^\mu U^{\nu}_c=2S^{\mu\rho}S_{\rho}^{\ \nu}+4S^{\mu\nu}-
2s(s+2)g^{\mu\nu}\tag5.10$$
This implies
$$P_cP+PP_c=2p_\mu p_\nu S^{\mu\rho}S_\rho^{\ \nu}-2s(s+2)p^2$$
In virtue of eq. \thetag{4.12}, which can be extended to quantum
theory without any change and also due to equation
$$S^{\mu\nu}S_{\mu\nu}=2U^2-2s(s+2)\tag5.11$$
we obtain
$$V^2+W^2+U^2p^2=-w^2\tag5.12$$
For $p^2=m^2\neq 0$ this relation is the quantum counterpart of eq.
\thetag{4.14}. For $m^2=0$ we get in analogy with \thetag{4.9}
$$V^2+W^2=w_1^2+w_2^2\tag5.13$$
Thus the generators $w_1,w_2,I_3,V,W,U$ form the Lie algebra
of  $E(2)\times E(2)$ group for which however the constraint
\thetag{5.13} holds.

We have come to the conclusion that the notion of AT has a well defined
physical and mathematical meaning. The object under study is a top/rotator,
since it is described by the group $SO(3)\times SO(3)$ of right - handed and
left - handed rotations characteristic for massive tops and similar group
$E(2)\times E(2) $ for massless tops/rotators. It is an Aristotelian
top/rotator since there is no term in the Routh function corresponding to
the rotational kinetic energy (see \thetag{3.8} and \thetag{7.2}) and so
there is no rotational excitation of the top/rotator. In quantum theory the
operator ordering can be always chosen so that the top has the only spin
value $ s = 1/2 $, in accordance with no excitation assumption.
\vskip \aaa
      \centerline{\bf 6. Quantum mechanical states}
\vskip \bbb
       In the beginning of Sec. 5 it was mentioned that by
fixing the operator ordering in the expressions for $\vek N,\vek U$ and
$\vek U_c$, which is controlled by a parameter $s$, one can
select spin $1/2$ particle as the only allowed state of the massive
top. On the first sight this conclusion may seem controversial
since intuitively one expects from a wave function of three variables
like $\Phi(\xi_i)$ the existence of infinite many states.
Therefore it will be instructive to repeat similar arguments and
as we shall see, with similar results for the massless top.

The wave function $\Phi(x,\xi_i)$ of a free massless
top must obey the equations
$$p^2\Phi=0\tag6.1$$
$$-(V^2+W^2)\Phi=\la\Phi\tag6.2$$
(the minus sign in \thetag{6.2} reflects the fact that the true
quantum mechanical operators are $-iV$ and $-iW$ ). The solutions
of \thetag{6.1} and \thetag{6.2} should be sought in the space, in which the
unitary representations of the group $E(2)\times E(2)$ are defined.
Since $E(2)$ is non-compact these representations are either
infinite-dimensional or one-dimensional. The former should be
excluded according to conventional wisdom \cite{13}, since otherwise they
would lead to infinite degeneracy
of states with a given momentum $p$. Such degeneracy would
cause e.g. inconsistencies in the description of statistical
properties of systems consisting of such particles. Therefore
one should restrict to one-dimensional representations, which
correspond to $\la=0$.

                From the point of view of $SO(3,3)$ group this restriction
corresponds to integral or half-integral $s$. The   value
$s=1/2$ defines an operator ordering which selects spin $1/2$ as
the only allowed spin value of the particle.

                 For a given value of $\la$ the wave function $\Phi$  is
specified also by eigenvalues of the operator $-iV$ or $-iW$.
Since $\la=0$ one obtains instead of \thetag{6.2} two equations
$$V\Phi=0$$
$$W\Phi=0\tag6.3$$
In \cite{7} matrix representations of V and W  for $s=1/2$ were found
$$V\Phi=\Omega^+V_{matr}\Psi$$
$$W\Phi=\Omega^{-}W_{matr}\Psi\tag6.5$$
where
$$V_{matr}=\frac{1}{2}\ga^\mu\ga_5p_\mu$$
$$W_{matr}=\frac{i}{2}\ga^\mu p_\mu\tag6.5$$
$$\ $$
$$\Phi(x,\xi_i)=\Omega^+(\xi_i)\Psi(x)$$
$$\Omega^+(\xi_i)=(\chi^+,\zeta^+)$$
$$\ $$
$$\chi^+=\xi^{-\frac{i}{2}}(ie^{i\frac{\fii}{2}}\sin{\frac{\theta}{2}},
            e^{-i\frac{\fii}{2}}\cos{\frac{\theta}{2}})$$
$$\zeta^+=\xi^{\frac{i}{2}}(e^{i\frac{\fii}{2}}\cos{\frac{\theta}{2}},
           ie^{-i\frac{\fii}{2}}\sin{\frac{\theta}{2}})\tag6.6$$
and $\Psi$ is Dirac wave function in the spinor representation.
Eq. \thetag{6.3} then imply
$$\ga^\mu p_\mu\Psi=0\tag6.7$$
           The conclusion, which can be drawn from these considerations
is quite remarkable: under the accepted operator
ordering the only quantum mechanical states of the massless
top are those, which satisfy eq. \thetag{6.7}. This is to be compared
with classical states, which are restricted by the condition
$p^2=0$ only, leaving $\xi_i,\eta_i$ arbitrary.

The explanation of this paradox will be discussed in Conclusions, where it
will be indicated that the restriction to finite dimensional representations
of Lorentz group has no real ground. In fact it is sufficient to choose a
weaker condition, namely that the spin should be $1/2$, but the
representation of Lorentz group may be both finite and infinite dimensional.
In this case $\lambda^2 $ is arbitrary and the quantum counterpart of the
classical AT is in general described by a wave function $\Phi (x, \xi_i)$
with the transformation properties, which correspond to the infinite
($\lambda^2 \neq 0 $) and finite ($\lambda^2 =0 $) dimensional representation 
of Lorentz group. 

\vskip \aaa
\centerline{\bf 7. Classical relativistic Aristotelian rotator}
\vskip \bbb

In this section we will give the relativistic version of the theory
considered in Sec. 3. Again we start with the action
  $$ S_R=\int_{\tau_1}^{\tau_2}(\eta_i \dot{\xi_i} - R)d\tau -\xi_{2i} 
                         \eta_{2i} \tag 7.1  $$
The evolution parameter $ \tau $ is associated with the worldline 
$ x^\mu=x^\mu(\tau) $ of the rotator and the Routh function $R$ is assumed 
to have the form
  $$ R= \frac{\mu}{2} u^2 + \frac{m^2}{2\mu} + ( eA_\mu + e_Z Z_\mu 
        + e_Z' UZ_\mu )u^\mu + \frac{1}{2\mu}(eF_{\mu\nu} + e_ZF_{\mu\nu}^Z +
                      e_Z'UF_{\mu\nu}^Z )S^{\mu\nu}  \tag 7.2 $$
where $ F_{\mu\nu}^z =\partial_\mu Z_\nu - \partial_\nu Z_\mu $ and
$u^{\mu}=\frac{dx^{\mu}}{d\tau}$. The action \thetag{7.1} is
reparametrization invariant, which is guaranteed by auxiliary variable $\mu$
transforming under the reparametrization in the following way
  $$ \tau '=f(\tau), \qquad \mu '=\frac{df(\tau)}{d\tau} \mu $$
This variable is determined by fixing the evolution parameter. The proper
time interpretation of $\tau$ corresponds to $\mu=m$, where $m$ is the mass
of the rotator.

There is no rotational kinetic energy term in \thetag{7.2}, rotator cannot
be rotationally excited, it is of Aristotelian type. Moreover the action
\thetag{7.1} is invariant under gauge transformations  
    $$ \align
       &\xi_i' = e^{-\Lambda} \xi_i      \\
       &\eta_i' = e^\Lambda \eta_i \tag 7.3  \\
       &Z_\mu' = Z_\mu - \frac{1}{e_Z^{'}} 
               \frac{\partial \Lambda}{\partial x^\mu}    \endalign   $$
which guarantees that the particle under consideration is really a rotator.
The action \thetag{7.1} is of course invariant also with respect to the
ordinary gauge transformations of electromagnetic potentials $A^{\mu}$. Note
that we have included in $R$ terms proportional to the coupling constant
$e_Z$ in order to obtain the most general coupling to $Z^{\mu}$. From the
quantum theory of the rotator based on the Dirac equation (see Sec. 9) it
follows that
  $$ e_Z'=\frac{\bar g}{2}; \qquad e_Z=\frac{\bar g}{4} -
               \frac{{g'}^2}{\bar g}; \qquad e=\frac{gg'}{\bar g} \qquad
       \text{where} \qquad \bar g = (g^2 + {g'}^2)^{\frac{1}{2}}  \tag 7.4 $$
and $g$, $g'$ are the conventional coupling constants of EWSM. The
complete action includes also the field contributions $S_A$, $S_Z$
  $$ S= S_R + S_A + S_Z $$
  $$ \align 
     &S_A = -\frac{1}{4} \int F_{\mu\nu} F^{\mu\nu} d^4x \\
     &S_Z = -\frac{1}{4} \int F^Z_{\mu\nu} F^{Z\mu\nu} d^4x  \endalign $$

The classical Hamiltonian which corresponds to a Routh function $R$ is
defined as follows
  $$ H =-p_{\mu} u^{\mu} +R  \tag 7.5  $$
where $p_{\mu} = \frac{\partial R}{ \partial u^{\mu}}$. With $R$ given by
$\thetag{7.2}$ this leads to
  $$ H=\frac{1}{2\mu} \lbrack m^2 - ( p_{\mu} - eA_\mu - e_Z Z_\mu 
        - e_Z' UZ_\mu )^2 + (eF_{\mu\nu} + e_Z F_{\mu\nu}^Z +
                      e_Z'UF_{\mu\nu}^Z )S^{\mu\nu} \rbrack \tag 7.6 $$ 
Variation of the action $\thetag{7.1}$ with respect to $\mu$ leads to the
condition
  $$ \frac{\partial R}{\partial \mu} = 0 $$
From $\thetag{7.6}$ it then follows
  $$ \frac{\partial H}{\partial \mu} =0  $$
or taking into account $\thetag{7.6}$
  $$ H= 0   \tag 7.7 $$
The canonical equations for orbital variables are
  $$ \align 
     &\frac{dx^{\mu}}{d\tau} = -\frac{\partial H}{\partial p_{\mu}} \\
     &\frac{dp^{\mu}}{d\tau} = \frac{\partial H}{\partial x_{\mu}} \tag 7.8 
                                                      \endalign  $$
As for the internal (rotational) variables the equations of motion can be
most simply written in the Poisson bracket form
  $$ \frac{dS^{AB}}{d\tau} = \lbrace H,S^{AB} \rbrace \tag 7.9 $$
The explicit form of this equation can be obtained using the Poisson
bracket relation $\thetag{4.1}$.      
\vskip \aaa
\centerline{\bf 8. Two kinds of rotators and tops}
\vskip \bbb
  In this section we will show that there are two kinds of rotators and tops
and the difference between them is manifested by the different behaviour under
Lorentz transformations.

 To this aim let us first introduce slightly generalized canonically conjugated
 4-component variables $ \xi_\mu = ( \xi_0 , \xi_i ) $, $\pi_\mu = 
( \pi_0 , \pi_i ) $, which are however subjected to constraints
$$ \align 
   \xi_0^2 - \xi_i^2 &= 0  \\
    \pi_0^2 - \pi_i^2 &= 0   \tag(8.1) \endalign $$
Due to these constraints Poisson brackets should be replaced by Dirac brackets 
$$ \align
  \{ \xi_\mu , \pi_\nu \} &= g_{\mu\nu} - \frac{\xi_\nu \pi_\mu}{\xi\pi}  \\
  \{ \xi_\mu , \xi_\nu \} &= \{ \pi_\mu , \pi_\nu \} = 0   \tag 8.2 
                                        \endalign         $$
where $ \xi\pi = \xi_0 \pi_0 -\xi_i \pi_i $ and $ g_{\mu\nu} $ is the usual 
space-time
metric tensor. One can easily verify that constraints $\thetag{8.1}$ are 
consistent with
Dirac bracket relations $\thetag{8.2}$ . Note that despite the vectorial form 
of eq. $\thetag{8.1}$ 
and $\thetag{8.2}$ $ \xi_\mu $ and $ \pi_\mu $ are not fourvectors! 

 The relation between $ \eta_i $ and $ \pi_\mu $ is 
$$ \eta_i = \pi_i - \frac {\pi_0}{\xi_0} \xi_i             \tag 8.3 $$
The inverse one reads 
$$ \pi_i= \eta_i -\frac {\eta_k^2}{2\xi_l \eta_l } \xi_i;\qquad \pi_0 =-\frac 
{\eta_k^2}{2\xi_l \eta_l } \xi_0  \tag 8.4 $$
 As for $ \xi_\mu $ the relation is simple
$$ \xi_\mu = ( \pm \xi , \xi_i )         \tag 8.5  $$
One can show that from the Dirac bracket relations $\thetag{8.2}$ and the eq.
$\thetag{8.3}$
the standard Poisson brackets relations follow 
$$ \align
      \{ \eta_i , \xi_k \} &= \delta_{ik}              \tag 8.6 \\
   \{ \eta_i , \eta_k \} &= \{ \xi_i , \xi_k \} = 0    \endalign $$
 as expected.

Note that 
$$ \pi_0 \dot\xi_0-\pi_i \dot\xi_i =-\eta_i \dot\xi_i \quad \text{and}\quad
   \pi_0 {\xi_0}-\pi_i {\xi_i}=-\xi_i {\eta_i} $$
 so that action integral $\thetag{7.1}$
can be easily expressed by means of variables $ \xi_\mu , \pi_\mu $. Also the
generators of the $SO(3,3)$ group can be expressed in new variables. First of 
all one gets from $\thetag{2.8}$
$$ \align
    s_1 &=\xi_2 \pi_3 - \xi_3 \pi_2 - \frac {\xi_0 \xi_1}{\xi_1^2+\xi_2^2}
    \xi \pi\\
    s_2 &=\xi_3 \pi_1 - \xi_1 \pi_3 - \frac {\xi_0 \xi_2}{\xi_1^2+\xi_2^2}
    \xi \pi
             \tag 8.7  \\ 
     s_3&=\xi_1 \pi_2 - \xi_2 \pi_1  \endalign $$  
Here we have used relations $\thetag{8.3}$ and the variables $ \xi = \sqrt{\xi_i
^2} $
 in $\thetag{2.8}$
was interpreted as $ \xi_0 $ in order that $ s_1$, $s_2 $ be single-valued 
functions of $\xi_\mu $.
This requirement is necessary to obtain unique extension of $ s_i $ from the 
region $ \xi_0 >0 $
to the region $ \xi_0 < 0 $. The expression $\thetag{8.7}$ for $ s_i $ are then 
linear functions
of $ \pi_\mu $ and rational functions of $ \xi_\mu $. In this way one can 
pass smoothly
from the value $\xi_0 = \sqrt{\xi_i^2} \equiv \xi $ to the value 
$ \xi_0 =-\xi $.
When one returns back to the variables $ \eta_i $ in $\thetag{8.7}$ one obtains 
$$ \align
  s_1&=\xi_2 \eta_3 - \xi_3 \eta_2 + \frac {\xi_0 \xi_1}{\xi_1^2+\xi_2^2} \xi_i
\eta_i \\
 s_2&=\xi_3 \eta_1 - \xi_1 \eta_3 + \frac {\xi_0 \xi_2}{\xi_1^2+\xi_2^2} \xi_i
\eta_i \tag 8.8 \\ 
 s_3&=\xi_1 \eta_2 - \xi_2 \eta_1  \endalign $$  
Note that $ s_i $ defined by ( 8.8 ) fulfil the standard Poisson brackets 
relations 
$$ \{ s_i,s_j \} = -\epsilon_{ijk}s_k \tag 8.9 $$
for both values of $ \xi_0=\pm \xi $. The same is true for all other 
generators of
$SO(3,3)$ group. To show that let us first generalize the unit vectors $
\vekgr{\nu}_3 $
and $ \vekgr{\nu} $ , originally defined for $ \xi_0 =\xi $ in the 
Appendix A. The generalized
expressions read 
$$  \align
\nu_1 &=-\frac{\xi^i}{\sqrt{\xi_1^2+\xi_2^2}} ( \frac{\xi_1 \xi_3}{\xi_0} +
                i\xi_2) \\
 \nu_2 &=-\frac{\xi^i}{\sqrt{\xi_1^2+\xi_2^2}} ( \frac{\xi_2 \xi_3}{\xi_0}
       -i\xi_1)   \tag 8.10 \\
 \nu_3 &=\frac{\xi^i}{\xi_0} {\sqrt{{\xi_1}^2+{\xi_2}^2}}  \\
 \nu_{3k} &=\frac{\xi_k}{\xi_0}        \endalign  $$
Actually, one can easily show that ${\vekgr{\nu}}_3$  and
$ \vekgr{\nu} $ obey the relations             
$$  \vekgr{\nu}^2 =1                   \hskip 1cm
    \vekgr{\nu}. \vekgr{\nu}^*=2       \hskip 1cm
    \vekgr{\nu}^2=0                    \hskip 1cm
    \vekgr{\nu}_3.\vekgr{\nu}=0        \tag 8.11$$
    $$  \vekgr{\nu}_3 \times \vekgr{\nu}=i\vekgr{\nu}           \hskip 1cm
        \vekgr{\nu} \times \vekgr{\nu}^* = 2i{\vekgr{\nu}_3} $$
irrespective of sgn $ \xi_0 $. Furthermore it holds 
$$ \align
 \{ s_i,\nu_{3j} \} &= -\epsilon_{ijk}\nu_{3k}           \tag 8.12 \\  
 \{ s_i,\nu_j \} &= -\epsilon_{ijk}\nu_k        \endalign $$          
 also for both signs of $ \xi_0 $. 

Taking into account the definitions of the remaining $SO(3,3)$ generators
$$ \align \vek N &= \vek s \times \vekgr {\nu}_3 \\
  U^0 = \vek s . \vekgr{\nu},&\qquad \vek U =i\vek s \times \vekgr{\nu}   
\tag 8.13 \\   
 U^0 = \vek s . {\vekgr{\nu}}^{*},&\qquad {\vek U}^* =-i\vek s
 \times \vekgr{\nu}^*  \\
   U &= \vek s . \vekgr {\nu}_3  \endalign $$
and eq. $\thetag{8.11}$, $\thetag{8.12}$, one sees that the Poisson brackets 
of any two generators
lead to the same result , regardless of the sign of $ \xi_0 $ and so 
the relations
$\thetag{4.1}$ hold for $ \xi_0 = \pm \xi $. Since the action of $SO(3,3)$ 
does not shift any
point from the region $ \xi_0 = \xi $ to $ \xi_0 = -\xi $ and vice versa, 
$ \xi_0 / \xi $
behaves with respect to $ S^{AB} $ like a constant and one can express this 
fact formally as follows 
$$ \{ S^{AB} , \frac {\xi_0}\xi \} =0     \tag 8.14 $$

 In conclusion one can say that there are two actions $\thetag{7.1}$ 
associated with two
different Routh functions, denoted as $ R_{u,d} $.
In $ R_u $ the generators $ S^{AB}$ correspond conventionally to 
$ \xi_0=-\xi$,
while in $R_d$ they correspond to $ \xi_0=+\xi $. Besides this the 
coupling constants:
 $e$, $e_Z$, $e_Z'$ are in general different in $R_u$ and $R_d$.
In the next section we shall argue that these two Routh functions correspond 
to two
fermions with different electric charges. The main argument will be based 
on the quantum version 
of eq. $\thetag{8.14}$ indicating that there is an observable 
$ (\xi_0 / \xi ) $ whose 
operator commutes with the operators of all other observables, associated with 
the system under consideration. 
Such an operator generates the superselection rule
connected with the electric charge.

Here we shall restrict ourselves to another, more formal, distinction 
between the tops (~ rotators ) corresponding 
to ${\xi }_{0}=\pm \xi $ as manifested by the transformation properties under 
Lorentz group. An infinitesimal Lorentz transformation is characterized by the 
generating function $ F= - \vek {N} \cdot  \delta \vek {V}
 - \vek {s} \cdot \delta \vekgr {\phi } $ and the 
corresponding transformation of ${\xi}_{i} $ reads $$ {\xi }_{i} \rightarrow 
{\xi }_{i}+ \{ F, {\xi}_{i} \}= {\xi}_{i}+ \frac{ \partial F }{ \partial {\eta}_
{i}}
 \tag 8.15 $$ 
Since  $\vek {N}, \vek {s} $ have different form for
${\xi }_{0}=\pm \xi $ the variables $ {\xi}_{i} $ transform differently 
in these two cases. 
In particular the unit vector ${\nu}_{3i}= \frac {{\xi}_{i}}{\xi} $
transforms under rotations like an ordinary three-vector and under Lorentz 
boosts according to a non-linear transformation law:$$ \frac{{\xi}_{i}}{\xi} 
\rightarrow \frac{{\xi}_{i}}{\xi}- \frac{{\xi}_{0}}{\xi}( \delta v_{i}-
                         \frac{{\xi}_{i} {\xi}_{j}}
{{\xi}^{2}} \delta v_{j}) \tag 8.16 $$ 
In either cases the quadratic form 
$ {\nu}_{3i}^{2} $ 
is invariant, so that also the equation 
$$ {\xi}_{0}^{2}-{\xi}_{i}^{2}=0 \tag 8.17 $$
is Lorentz invariant. In the special case of rotator the vector 
$\vekgr {\nu}_3 $ 
determines its symmetry axis and the eq. $\thetag{8.16}$ gives 
two 
different transformation rules for this axis under Lorentz boosts. 

\vskip \aaa
\centerline{\bf 9. In the search of a rational reconstruction of EWSM}
\vskip \bbb

   In this and the next Section we will summarize and analyze the logical
steps necessary for a final reconstruction of EWSM from the string dynamics
via the Aristotelian top. In the first step we must realize that the string
dynamics at low energies is greatly influenced by the large energy value of
the first excitation level $O(E_{Pl})$. Consequently for energies $E \ll
E_{Pl}$, which cover the region of validity of EWSM the vibrational and
rotational degrees of freedom cannot be excited, they are frozen. This
however does not mean that there cannot be any rotational motion at all. We
assume that (besides the orbital motion of the string as a whole) a
rotation in the external field is possible, leading to the process of spin
precession. Such an assumption was tested in [6] and it was shown there that
it leads to consistent theory of spinning particle. We also anticipate that
in the Higgs field string can change its shape, causing in this manner the
gauge symmetry breaking, without generating the vibrational excitations. We
thus come to the conclusion that an Aristotelian regime sets up when we
approach the low energy limit. In this regime the motion of string (rotation
and shape deformation) is possible only if the string is subjected to some
external cause - an external (gauge or/and Higgs) field. 

   As for the equation of motion for the Aristotelian string one can
extend the theory  elaborated in (6) for motion in an electromagnetic field
to any gauge field. This was done in previous Sections. However we have no
clear idea how to get the interaction with Higgs field from the first
principles. In such situation one can borrow the whole Higgs sector from
EWSM and assume that string is straight line - a rotator - in the false
Higgs vacuum. The bending of string is associated with the transition of
Higgs field to the true vacuum. However one does not see how the curvature
of string is controlled by the Higgs field in such transition. From this
point of view a more appealing approach is offered by the theory of
elasticity, where the phenomenon of elastic instability, similar to that
generated by Higgs field, is known for long time. It consists in bending of
beam, whose one end is fixed and the other is loaded by a force acting
parallel to the beam. It is known that when the force exceeds some critical
value the initially straight beam bends and the axial rotational symmetry is
violated. Note that this process is given in [14] as an example of
spontaneous symmetry breaking in connection with EWSM!

   In the present paper we will not develop these speculative ideas further
and in what follows in constructing the Aristotelian equations of motion we
will restrict ourselves to gauge fields and to the rotational degrees of
freedom, only. The interaction with the Higgs field will be taken over from
EWSM.

   In the second step we will take into account the consequences resulting
from the fact that there is no term in the Hamiltonian $\thetag{7.6}$
corresponding to the kinetic energy of the rotational motion. Due to this
there is no relation between the angular velocity $\dot{\xi}_i$ and the
canonical momenta $\eta_i$, and in virtue of this there is no quantum
mechanical correlation between $\xi_i$ and $\eta_i$. In the conventional
quantum theory of point particles the relation $p_i=mx_i$ guarantees the
expansion of wave packet and equivalence of canonical and path integral
quantization. When the rotational kinetic energy is missing in the
Hamiltonian the path integration does not lead to any expansion of spin wave
packet and the equivalence between the two procedures of quantization is
lost.

   On the first sight this may signal an internal inconsistency of AT, but a
closer look reveals that it is not so. In Sec. 2 and 3 we have seen that the
rotator mode of the top can be realized only through the interaction with
some gauge vector field which is responsible for the rotator $U(1)$
symmetry. Thus, there cannot be free rotator/top, it must interact with the
gauge field. At least it must be thought as interacting with the vacuum
fluctuations of the gauge fields. This interaction generates the quantum
mechanical dispersions of the Euler angles, or equivalently $\xi_i$. In
other words it leads to the expansion of the $\xi_i$-part of the wave
function. To see that let us consider a propagation of AT from the point
$x_1$ to $x_2$ and let us assume that the variables $\xi_i$ have in $x_1$
sharp values $\xi_i=\xi_{1i}$. The path connecting $x_1$ and the $x_2$ goes
through the gauge field, so that $\xi_i$ is changed depending on the path.    
In $x_2$ the wave function will then be in 
general different from zero for all
possible values of $\xi_i$. In [7] the propagator of AT in an external
electromagnetic field was derived by means of path integrals and it was
shown there that the spin wave functions $\chi^+$, $\zeta^+$ given by
$\thetag{6.6}$ are reproduced in the propagation of AT. As a result any spin
state can be expressed as linear combination of the components of $\chi^+$,
$\zeta^+$. But these functions were in Sec. 6 introduced as the result of
canonical quntization, so that there is a strong evidence that canonical and
path intergral quantization can be reconciled if one takes AT as whole, i.e.
including its gauge fields.

   If so, one can perform in the third step canonical quantization by formal
replacement $\eta_i \longrightarrow \frac{\partial}{\partial \xi_i}$, as
shown in Sec. 5. By means of a suitable ordering of operators in the
expressions for $S^{AB}$ one can achieve that $s=\frac{1}{2}$ is the only
allowed
value of spin of the top, in accordance with its Aristotelian nature. The
representation space of the generators $S^{AB}$ is spanned on the basis
$D^{1/2}_{km}(\vartheta, \varphi, \psi)$, $k=\pm \frac{1}{2}$, $m=\pm
\frac{1}{2}$, denoted here as $\chi^+$, $\zeta^+$ (see $\thetag{6.6}$). A
massless top is described by the group $E(2)\times E(2)$ (Sec. 5). Its wave
function $\varPhi$ fulfils the equation $\thetag{6.3}$
 $$ V\Phi =0, \qquad W\Phi =0 \tag 9.1 $$
where $V$ and $W$ are two commuting generators of the group $E(2)$. The
solution of $\thetag{9.1}$ can be expressed as
$\Phi=\Omega^+(\xi_i)\Psi(x)$ where $\Omega^+=(\chi^+,\zeta^+)$ and
$\Psi$ is the solution of Dirac equation (see Sec. 6). 

\vskip \aaa
\centerline{\bf 10. An attempt to derive $SU(2)_L\times U(1)_Y$ gauge symmetry}
\vskip \bbb

In the forth step on the basis of Sec. 8 we take first into account two kinds
of
rotators (tops). This allows us to distinguish between up type and down type
fermions and to introduce the weak isospin. The configuration space consists
of two subspaces parametrized by the coordinates $\xi_i$, $\xi_0=\pm
\sqrt{\xi_i^2}$. The representation space is spanned on the basis
$\th(\xi_0)\Omega^+(\xi_i)$, $\th_0(-\xi_0){\Omega '}^+(\xi_i)$,
where $\Omega^+(\xi_i)$
is defined by (6.6) and ${\Omega '}^+=({\chi '}^+,{\zee '}^+)$ forms the
spinorials
basis for $\xi_0=-\xi$. To find this basis we first observe that the transition
from $\xi_0=\xi$ to $\xi_0=-\xi$ in the generators $S^{AB}$ is equivalent
to the transformation $\xi_i\to-\xi_i$, $\eta_i\to -\eta_i$ (see eq. (8.8)).
In case of generators $U^\mu$, $U^\mu_c$, in addition, one has to change
the sign: $U^\mu\to-U^\mu$, $U^\mu_c\to-U^\mu_c$, too (see the expressions
(8.10) for $\nu_i$). However by a redefinition of $U^\mu$, $U^\mu_c$ one
can omit this change of sign, so that the transition to region 
$\xi_0\to-\xi$ is then the same for all $S^{AB}$ and is realized through
the replacement $\xi_i\to-\xi_i$. Note that this redefinition is consistent
with the commutation relations for $S^{AB}$ (5.1). 

The basis for $\xi_0\to-\xi$ is then defined as follows
$${\zee '}^+=\zee^+(-\xi_i)=\xi^{\frac{i}{2}}(ie^{\frac{i}{2}\varphi}
                        \sin{\frac{\th}{2}},e^{-\frac{i}{2}\varphi}
                        \cos{\frac{\th}{2}})$$
$${\chi '}^+=\chi^+(-\xi_i)=\xi^{-\frac{i}{2}}(-ie^{\frac{i}{2}var\phi}
                        \cos{\frac{\th}{2}},-ie^{-\frac{i}{2}\varphi}
                        \sin{\frac{\th}{2}})\tag10.1$$
The extended generators have the form
$$S^{AB}_{\text{ext}}=\th(\xi_0)S^{AB}(\xi_i)+\th(-\xi_0)S^{AB}(-\xi_i)
\tag10.2$$
and the corresponding wave functions are
$$\varPhi_{\text{ext}}(x,\zee_\mu)=\Omega^+(\xi_i)e(x)\th(\xi_0)+
                           \Omega^+(-\xi_i)\nu (x)\th(-\xi_0)\tag10.3$$
where $e(x)$ and $\nu (x)$ are the wave functions of charged and neutral
leptons respectively.

As indicated in Sec. 8, a superselection rule holds in the 
representation space of the operators $S^{AB}_{\text{ext}}$. The operator
$\frac{\xi_0}{\xi}$ commutes with all $S^{AB}$  
$$[S^{AB}_{\text{ext}},\frac{\xi_0}{\xi}]=0\tag10.4$$
and the relative phases of functions $\Omega^+(\xi_i)e(x)\th(\xi_0)$
and $\Omega^+(-\xi_i)v(x)\th(-\xi_0)$ are arbitrary, since their product
is always zero due to $\th(-\xi_0)\th(\xi_0)=0$. As a result
the representation space can be decomposed into the
direct sum of two incoherent subspaces labeled by two
discrete values of
$\frac{\xi_0}{\xi}$. The only interpretation of this superselection
rule that can be considered in the present context is associated
with the electric charge and this is the reason why the states
belonging to different $\frac{\xi_0}{\xi}$ have been interpreted as up
and down type fermions. With respect to this duplication of fermionic
states the wave equations (9.1) must be replaced by
$$V_{\text{ext}}\varPhi_{\text{ext}}=0,\qquad W_{\text{ext}}\varPhi_{\text{ext}}=0\tag10.5$$
where $V_{\text{ext}}$ and $W_{\text{ext}}$ are defined by means of
$S^{AB}_{\text{ext}}$.

The equations (10.5) as they stand hold for genuine tops. To extend
their validity to rotators one must implement the axial symmetry of
rotators.

This leads to next - the fifth - step, the introduction of the gauge
symmetry of rotator. So far we have considered only symmetry
$U(1)_A\times U(1)_Z$, because in Sec. 3 and 7 we have dealt with
classical rotators, for which the transitions between the configuration
subspaces with $\frac{\xi_0}{\xi} =\pm 1$ cannot occur.
In quantum
theory such transitions are allowed (by emission and absorption
of $W^{\pm}$ bosons) and the definition of the
rotator gauge symmetry must be generalized correspondingly. To this aim
we first introduce the generators of the weak isospin
$$T_1=\frac{1}{2}I,\qquad T_2=\frac{i}{2}\frac{\xi_0}{\xi}I,\qquad
T_3=-\frac{1}{2}\frac{\xi_0}{\xi} \tag 10.6$$
where $I$ is an inversion operator defined in the space of functions
(10.3) as follows
$$I\th(\xi_0)=\th(-\xi_0),\qquad I\Omega^+(\xi_i)=\Omega^+(-\xi_i)$$
One can easily show that
$$T_iT_k+T_kT_i=\frac{1}{2}\de_{ik}\tag10.7$$
To make the derivation of $SU(2)_L\times U(1)_Y$ gauge symmetry
more transparent, especially to show that it has its roots in the axial
symmetry of rotator, we start with the simplest situation:

$i)$ Two neutral gauge fields $A_\mu$, $Z_\mu$ and one kind of rotator.
In the classical case we have seen in Sec. 3 and 7 that the symmetry is
$U(1)_A\times U(1)_Z$ with the generators 1 and $U+U_0$, where
$U=\vekgr\nu_3.\vek s$ is the projection of spin to symmetry axis and
$U_0$ is some constant. In quantum theory the second generator becomes
$-iU+U_0$, where $U=\vekgr\nu_3.\vek I=\xi_i\frac{\partial}{\partial
\xi_i}=\xi\frac{\partial}{\partial\xi}$ (see eq. (5.5))). Since the 
projection of spin can assume only two values $s_3^,=\pm 1/2$ it is
useful to introduce new gauge symmetry generators
$$P_{L,R}=\frac{1}{2}\mp iU\tag10.8$$
which are the projection operators to states with left-handed and 
right-handed chirality
$$P_L\Omega^+\psi=\frac{1}{2}\Omega^+(1+\ga_5)\psi=\Omega^+\psi_L$$
$$P_R\Omega^+\psi=\frac{1}{2}\Omega^+(1-\ga_5)\psi=\Omega^+\psi_R$$
where
$$\ga_5=\pmatrix -1&0\\0&1\endpmatrix$$
The gauge fields belonging to generators $P_L$ and $P_R$ will be
denoted as $W_{3\mu}$ and $B_\mu$, respectively. The equations (6.3)
in the explicit transcription read
$$S^{\mu 4}p_\mu\varPhi=0,\qquad S^{\mu 5}p_\mu\varPhi\tag10.9$$
The interaction corresponding to rotator mode is introduced by the 
replacement
$$p_\mu=i\partial_\mu\to i\partial_\mu+\frac{g}{2}P_L W_{3\mu}
+\frac{g^,}{2}P_RB_\mu\tag10.10$$
where $g$ and $g^,$ are the coupling constants. The gauge symmetry is
then $U(1)_L\times U(1)_R$. The relations to the original fields are
given by the conventional rotation
$$W_{3\mu}=\cos\th_W Z_\mu+\sin\th_W A_\mu$$
$$B_{\mu}=-\sin\th_W Z_\mu+\cos\th_W A_\mu$$
where $\cos\th_W=\frac{g}{\bar g}$, $\sin\th_W=\frac{g'}{\bar g}$,
$\bar g=\sqrt{g^2+{g'}^2}$. From (10.10) one obtains
$$i\partial_\mu\to i\partial_\mu-eA_\mu+\frac{\bar g}{2}(U_0-iU)Z_\mu
\tag10.11$$
where $e=-\frac{gg'}{2\bar g}$ is the electric charge and $U_0=
\frac{1}{2}\cos 2\th_W$.

$ii)$ Two neutral gauge fields and both kinds of rotators. If we for a
while assume that the two kinds of rotators have the opposite electric
charge the generators of gauge transformation have slightly different
form: $\frac{\xi_0}{\xi}P_L$, $\frac{\xi_0}{\xi}P_R$. Instead of (10.10)
we will have
$$ i\partial_\mu \longrightarrow i\partial_\mu + \frac{g}{2}
        \frac{\xi_0}{\xi} P_L W_{3\mu} + \frac{g'}{2}\frac{\xi_0}{\xi} P_R
        B_\mu                                   \tag 10.12             $$       
and this replacement should be done in eq. $\thetag{10.5}$. The electric
charge is then $eQ$, where $e=\frac{gg'}{\bar g}$ and
$Q=-\frac{\xi_0}{2\xi}$. In fact the charges of up type and down type
fermions do not differ in sign only. Rather the charge of the whole doublet
is shifted by amount $Q_0$, but this can be without problems incorporated in
our gauge symmetry pattern, since each $U(1)$ symmetry generator is fixed up
to an additive constant. Taking into account that $\frac{1}{2}
\frac{\xi_0}{\xi}=-T_3$ we see that the gauge symmetry contained in
$\thetag{10.12}$ is $U(1)_{T_3 L} \times U(1)_{T_3 R}$. This leads us to the
final situation with \newline
$iii)$ two neutral and two charged gauge fields as well as two kinds of
rotators. From the comparison with EWSM we conclude that the final fixation
of gauge generators must be the following
$$ -T_3 P_L \longrightarrow -T_i P_L                   \tag 10.13  $$
$$ \frac{\xi_0}{\xi} P_R \cases \longrightarrow & 
         \frac{\xi_0}{\xi} P_R +1=-Y, \text{for leptons} \\
                                \longrightarrow &
         \frac{\xi_0}{\xi} P_R - \frac{1}{3}=-Y, \text{for quarks}    \endcases
                                                        \tag 10.14 $$ 
The replacement $\thetag{10.13}$ can be interpreted in our opinion as the
generalization of the rotator axial symmetry to the interaction with charged
gauge fields, besides the neutral ones. In fact the gauge symmetry has the
group character and the only generalization of $U(1)_{T_3}$ is $SU(2)$.

  As for replacement $\thetag{10.14}$ the problem is how to explain the
values of additive constants in $Y$. But the same problem was encountered in
EWSM, where it was solved using the condition following from cancellation of
triangular chiral gauge anomalies. In particular one gets for $Y$ the
condition
$$ \sum_{\text{leptons}}Y + \sum_{\text{quarks}}Y =0 \tag 10.15     $$
Since $\sum \frac{\xi_0}{\xi} =0$ it follows from $\thetag{10.14}$ that
condition $\thetag{10.15}$ is satisfied ($4.1 -3.4.\frac{1}{3} =0$).

  Taking the replacements $\thetag{10.13}$ and $\thetag{10.14}$ as at least
intuitively justified we obtain from $\thetag{10.12}$
$$ i\partial_\mu \longrightarrow i\partial_\mu 
            -gT_i P_L W_{i\mu} - \frac{g'}{2} Y B_{\mu}                $$
Inserting this in $\thetag{10.5}$ we get 
$$ S_{\text{ext}}^{\mu4} (i\partial_\mu 
            -gT_i P_L W_{i\mu} - \frac{g'}{2} Y B_{\mu})
            \varPhi_{\text{ext}} =0                    \tag 10.16      $$
$$ S_{\text{ext}}^{\mu5} (i\partial_\mu 
            -gT_i P_L W_{i\mu} - \frac{g'}{2} Y B_{\mu})
            \varPhi_{\text{ext}} =0                    \tag 10.17      $$
    To pass from differential operators to Dirac matrices we use the
relations found in [7]
$$ S^{AB}(\xi_i)\Omega^+(\xi_i)=\Omega^+(\xi_i)S^{AB}_{\text{matr}}
                                                        \tag 10.18     $$
where
$$ \align
  S^{\mu\nu}_{\text{matr}} &=- \frac{1}{4}(\gamma^{\mu}\gamma^{\nu} -
                  \gamma{\nu}\gamma^{\mu})    \\
  S^{\mu4}_{\text{matr}} &=- \frac{1}{2}\gamma^{\mu}\gamma_5 \\
  S^{\mu5}_{\text{matr}} &= \frac{i}{2}\gamma^{\mu} \\
  S^{45}_{\text{matr}} &= \frac{i}{2}\gamma_5  
                                                \endalign              $$
and the $\gamma$-matrices are given in the spinor representation
$$ \gamma^0 =\left(\matrix 0&1\\1&0 \endmatrix\right), \qquad
   \gamma^i =\left(\matrix 0&-\sigma_i\\\sigma_i&0 \endmatrix\right)   $$
The relations $\thetag{10.18}$ naturally hold also for $\xi_i
\longrightarrow -\xi_i$, so that $S^{AB}_{\text{ext}}$ can be expressed
by means of $\gamma$-matrices, too. Note that $\gamma$-matrices do
not provide room for the distinction between up and down type fermions,
while the differential operators $S^{AB}_{\text{ext}}$ such a room do
provide. From the expressions for $S^{\mu4}_{\text{matr}}$ and
$S^{\mu5}_{\text{matr}}$ it follows that the equations $\thetag{10.16}$ and
$\thetag{10.17}$ are not independent, so we restrict ourselves to
$\thetag{10.17}$ only.

   It is suitable to introduce
$$ \Omega^+_{\text{ext}} (\xi_0,\xi_i) =
          (\vartheta (-\xi_0)\Omega(-\xi_i),
              \vartheta (\xi_0)\Omega^+ (\xi_i))                       $$
Then 
$$ T_k \Omega^+_{\text{ext}} =\frac{1}{2} \Omega^+_{\text{ext}}\tau_k
                                                                        $$
where $\tau_k$ are weak isospin Pauli matrices. The conventional equations
for fermionic fields of EWSM follow from $\thetag{10.17}$ if we insert
$$ \varPhi_{\text{ext}} =   
    \Omega^+_{\text{ext}} \psi_L  +
          \vartheta(-\xi_0)\Omega(-\xi_i)\nu_R +
              \vartheta(\xi_0)\Omega^+(\xi_i)e_R                     $$
where
$$ \psi_L =\left(\matrix \nu_L\\e_L \endmatrix\right)                    $$
We obtain
$$\gamma^{\mu}(i\partial_\mu - \frac{g}{2}
        \vekgr{\tau}.\vek W_{\mu} + \frac{g'}{2} B_\mu)\psi_L=0,
        \qquad (h_e e_R \varphi + h_\nu \nu_R \varphi_c )               $$
$$\gamma^{\mu}(i\partial_\mu + g' B_\mu)e_R=0,
        \qquad (h_e \varphi^x \psi )                                    $$
$$\gamma^{\mu} i\partial_\mu \nu_R=0,
        \qquad (h_e \varphi^x_c \psi )                                  $$
where
$$ \varphi= \left(\matrix \varphi^+\\\varphi^0 \endmatrix\right), \qquad
   \varphi_c= \left(\matrix {\bar \varphi}^0 \\-\varphi^- \endmatrix\right)  $$
$$ \varphi^x= \left(\matrix \varphi^-,&\bar{\varphi^0} \endmatrix\right),
   \qquad
   \varphi_c^x= \left(\matrix \varphi^0,&-\varphi^+ \endmatrix\right)  $$
and $h_e$, $h_\nu$ are the coupling constants. In the brackets on the r.h.s.
of last equations are the terms with Higgs fields. These terms do not
follow so far from the model presented here and have been included for
completeness.                                        

\vskip \aaa
\centerline{\bf 11. Conclusions}
\vskip \bbb
The notion of AT introduced in this paper plays a useful mediatory role in
establishing the link between strings and the spin 1/2 particles. On one
hand this notion represents a suitable model of Dirac particles in the sense
that it allows only minimal value of spin $s=1/2 $ for the particle. On the
other hand AT can be interpreted as a string at low energies, when
vibrational and rotational degrees of freedom are frozen. AT offers also
some sort of reconciliation of two opposite interpretations of spin: one as
coming from the rotation of a rigid body and the other as a mere quantum
effect. According to the novel interpretation, spin is associated with the
rigid body, but this body is of Aristotelian nature, so that a free particle
does not rotate, although it has spin.

The mediatory role of AT has been exploited in the present paper for a
derivation of EWSM from the AT dynamics. The starting point was a peculiar
property of AT consisting in the fact that the rotator mode of AT can be
realized only through the gauge interaction with a vectorial field with the
gauge group $U(1)$. This group represents the axial symmetry of the rotator,
i.e. the symmetry under the rotations around the rotator axis. The second
fact relevant for the derivation of EWSM was the observation that there are
two kinds of rotators/tops. Arguments based on the superselection rule
indicate that they can be interpreted as up - type and down - type fermions.
In view of two types of rotators intuitive arguments lead to the conclusion
that in general the axial symmetry of the rotator can be realized through
the group $SU(2)\times U(1) $ and for the group $SU(2)$ one selects either
singlet or doublet representation (the former for right - handed, the latter
for left - handed fields).

Note some analogy with Kaluza - Klein interpretation of electromagnetic
$U(1)$ gauge symmetry in $5D$ classical field theory. The extra space
dimension is compacitfied on a circle, which possesses $U(1)$ isometry and
due to the general covariance this isometry is converted into $U(1)$
electromagnetic gauge symmetry. Here the isometry of the circle is replaced
by the rotator axial symmetry and the locality of gauge transformation is a
consequence of spin 1/2 of the rotator (Sec. 4).

If our representation of electroweak gauge symmetry is correct, then the
spontaneous violation of the symmetry must be associated with the bending of
rotator under the influence of Higgs field and subsequent transformation of
rotator to top. We have not studied in more details the models for such a
transformation, because it is a problem per se, obviously related with the
complicated issue of lepton and quark masses.

Instead, the problem which must be studied next is the $SU(3)$ color gauge
interaction of quarks. If our basic philosophy is correct, AT must include
not only leptons, but also quarks. Consequently in mathematical description
of AT there must be a room, which allows to introduce quarks as a new
particular sort of AT, in a similar way as it has been done in case of two
categories of fermions, the up - type and down - type ones. Preliminary
search in this direction revealed an interesting possibility, namely the
infinite dimensional representations of Lorentz and $SO(3,3)$ group. As
stressed several times before the spin of AT must be 1/2 and following the
conventional wisdom this requirement alone leads to Dirac equation and thus
to finite dimensional representation of Lorentz group. But this conclusion
is incorrect. Actually, spin is associated with the dimensionality of the
representation of the rotational group $SO(3)$ in the rest system of the
particle. Fixing the $s= 1/2 $ option we have still the liberty to select
either finite or infinite dimensional representation for the Lorentz boost.
Choosing the latter possibility we obtain a particle with properties in some
respects different from those, which share Dirac particles. In progress is a
study with the aim to find out whether these properties fit quarks or not.      

\vskip \aaa
\centerline{\bf Appendix A}
\vskip \bbb

For the sake of completeness we summarize here some kinematical definitions as 
well
as the definitions of SO(3,3) group generators $ S^{AB} $. We also present some
useful relations for $ S^{AB} $. The body coordinate system of the rotator/top 
is defined by means of unit vectors
$$ \vekgr {\nu}_1=( \cos \psi \cos \varphi - \sin \psi \sin \varphi \cos
\vartheta ,
\cos \psi \sin \varphi + \sin \psi \cos \varphi \cos \vartheta , \sin \psi \sin
\vartheta ) $$
$$ \vekgr {\nu}_{2}=( -\sin \psi \cos \varphi- \cos \psi \sin \varphi \cos
\vartheta,
-\sin \psi \sin \varphi+ \cos \psi \cos \varphi \cos \vartheta, \cos \psi \sin 
\vartheta) $$
$$ \vekgr {\nu}_{3}=( \sin \vartheta \sin \varphi,- \sin \vartheta \cos \varphi
, 
\cos \vartheta)  \tag A1 $$
where $ \vartheta, \varphi, \psi $ are Euler angles. A state of rotator/top is
characterized by the canonically conjugated variables $ {\xi}_{i}, {\eta}_{i} $,
 where $ {\xi}_{i} $ are related to Euler angles
$$ {\xi}_{1}=e^{\psi} \sin \vartheta \sin \varphi $$
$$ {\xi}_{2}=-e^{\psi} \sin \vartheta \cos \varphi $$
$$ {\xi}_{3}=e^{\psi} \cos \vartheta \tag A2 $$
and $ {\eta}_{i} $ to spin components
$$ s_{1}={\xi}_{2}{\eta}_{3}-{\xi}_{3}{\eta}_{2}+\xi 
\frac{{\xi}_{1}}{{\xi}_{1}^{2}+{\xi}_{2}^{2}}
{\xi}_{i}{\eta}_{i} $$
$$ s_{2}={\xi}_{3}{\eta}_{1}-{\xi}_{1}{\eta}_{3}+\xi 
\frac{{\xi}_{2}}{{\xi}_{1}^{2}+{\xi}_{2}^{2}}
{\xi}_{i}{\eta}_{i} \tag A3 $$
$$ s_{3}={\xi}_{1}{\eta}_{2}-{\xi}_{2}{\eta}_{1} $$
From the basic Poisson brackets relations
$$ \{ {\eta}_{i},{\xi}_{k} \}={\delta}_{ik} $$ 
$$ \{ {\xi}_{i},{\xi}_{k} \}= \{ {\eta}_{i},{\eta}_{k} \} =0 $$ 
one can derive the relations $$ \{ s_{i},s_{k} \}=-\varepsilon_{ikj}s_{j} $$
The generators $ S^{AB}=-S^{BA} $ are defined as follows
$$ (S^{23},S^{31},S^{12})=(s_{1},s_{2},s_{3}) $$
$$ (S^{01},S^{02},S^{03})=(N_{1},N_{2},N_{3}) $$
where $ \vek{N}=\vek{s} \times \vekgr {\nu}_{3}$. Furthermore
$$ S^{\mu 4}=- \frac{1}{2}(U^{\mu}+U^{\mu*}) $$
$$ S^{\mu 5}= \frac{1}{2i}(U^{\mu}-U^{\mu*}) $$
where $ U^{0}= \vekgr{\nu} \cdot \vek{s}, \vek{U}=-i \vekgr{\nu} \times \vek{s},
\vekgr{\nu}=\vekgr {\nu}_2+i \vekgr {\nu}_{1} $
The vectors $ \vekgr{\nu} $ and $ \vekgr {\nu}_{3} $ can be expressed by means
of 
$ {\xi}_{i} $:
$$ {\nu}_{1}=- \frac{{\xi}^{i}}{ \sqrt{{\xi}_{1}^{2}+{\xi}_{2}^{2} }}({\xi}_{1}
\frac{{\xi}_{3}}{\xi}+i{\xi}_{2}) $$
$$ {\nu}_{2}=- \frac{{\xi}^{i}}{ \sqrt{{\xi}_{1}^{2}+{\xi}_{2}^{2} }}({\xi}_{2}
\frac{{\xi}_{3}}{\xi}-i{\xi}_{1})  \tag A4 $$
$$ {\nu}_{3}= \frac{{\xi}^{i}}{\xi} \sqrt{{\xi}_{1}^{2}+{\xi}_{2}^{2} } $$
$$ {\nu}_{3i}= \frac{{\xi}^{i}}{\xi} , \xi= \sqrt{{\xi}_{i}^{2}} $$
Finally
$$S^{45}=U=\vekgr \nu_3 .\vek s =\xi_i \eta_i$$
The $SO(3,3)$ Lie algebra is given by the Poisson brackets relations
$$\{ S^{AB},S^{CD}\}= g^{BD}S^{AC}+
                      g^{AC}S^{BD}-
                      g^{BC}S^{AD}-
                      g^{AD}S^{BC} \tag A5$$
where
$$g^{00}=g^{44}=g^{55}=1, \hskip .5cm g^{11}=g^{22}=g^{33}=-1, \hskip .5cm
  g^{AB}=0 \ \text{for} \ A\neq B.$$
The following relations are direct consequences of the corresponding
definitions
$$S_{\mu \nu}S^{\mu \nu}=2U^2$$
$$\tilde S^{\mu \nu}S_{\mu \nu}=0$$
$$U_{\mu}U^{\mu}=0$$
$$U^{\mu *}U^{\nu} -U^{\nu *}U^{\mu}= -2iUS^{\mu \nu} \tag A6$$
$$U^{\mu *}U^{\nu} +U^{\nu *}U^{\mu}= 2S^{\mu \rho} S_{\rho}{}^{\nu} $$
$$S^{\mu \nu}U_{\nu} = iUU^{\mu}$$
where
$$\tilde S^{\mu \nu }= \frac 12 \varepsilon^{\mu \nu \rho \sigma} S_{\rho
\sigma}$$
Canonical quantization consists in the replacement $\eta_i\to
\frac{\partial}{\partial \xi_i}$. From $s_i$ we then obtain
$$I_1=\xi_2 \frac{\partial}{\partial \xi_3} -
      \xi_3 \frac{\partial}{\partial \xi_2} +
      \xi
      \frac{\xi_1}{\xi_1^2+\xi_2^2} \xi_i \frac{\partial}{\partial \xi_i}$$
$$I_2=\xi_3 \frac{\partial}{\partial \xi_1} -
      \xi_1 \frac{\partial}{\partial \xi_3} +
      \xi
      \frac{\xi_2}{\xi_1^2+\xi_2^2} \xi_i \frac{\partial}{\partial \xi_i}
                                                                 \tag A7$$
$$I_3=\xi_1 \frac{\partial}{\partial \xi_2} -
      \xi_2 \frac{\partial}{\partial \xi_1}$$
Other $S^{AB}$ generators are
$$\vek N = - \vekgr \nu_3 \times \vek I +s \vekgr \nu_3$$
$$U^0 = - \vekgr \nu . \vek I $$
$$\vek U = - i \vekgr \nu \times \vek I +is \vekgr \nu$$
$$U =  \vekgr \nu_3 . \vek I = \xi_i \frac{\partial}{\partial \xi_i}$$
where $s$ is an arbitrary constant. Additive terms in $\vek N$ and $\vek U$
are due to ordering ambiguity ($\vek I$ does not commute with $\vekgr
\nu_3$ and $\vekgr \nu$). $U^0_c, \vek U_c$ are complex conjugates of
$U^0, \vek U$. With the same definitions of $S^{AB}$ as in the
classical case we obtain
$$[ S^{AB},S^{CD}]= g^{BD}S^{AC}+
                      g^{AC}S^{BD}-
                      g^{BC}S^{AD}-
                      g^{AD}S^{BC} \tag A8$$
The quantum analogues of \thetag{A6} are
$$S_{\mu \nu}S^{\mu \nu}=2U^2 -2s(s+2)$$
$$\tilde S^{\mu \nu}S_{\mu \nu}=-4(1+s)U$$
$$U_{\mu}U^{\mu}=0$$
$$\frac{1}{2i}
  (U^{\mu}_cU^{\nu} -U^{\mu}U^{\nu}_c)= -US^{\mu \nu} +(1+s)\tilde S^{\mu \nu}
  -g^{\mu \nu}U \tag A9$$
$$\frac 12
  (U^{\mu}_cU^{\nu} +U^{\mu}U^{\nu}_c)= S^{\mu \rho} S_{\rho}{}^{\nu}
  +2S^{\mu \nu} - s(s+2)g^{\mu \nu}$$
$$S^{\mu \nu}U_{\nu} = i(U+i)U^{\mu}$$

\vskip \aaa
\centerline{\bf Appendix B}
\vskip \bbb

Conventional configuration space of strings is a set of $3D$ curves given by
the parametric equation $\vek{x} = \vek{x} (\sigma) $, where $\sigma$ is the
length along the string. By means of Frenet's equations
$$ \align
   \frac{d\vek{t}}{d\sigma} &= k \vek{n}\\
    \frac{d\vek{n}}{d\sigma} &= -k \vek{t} + \varkappa \vek{b} \tag B1 \\
  \frac{d\vek{b}}{d\sigma} &= -\varkappa \vek{n} \endalign $$
one can obtain two other configuration spaces. Here $\vek{t}$, $\vek{n}$,
$\vek{b}$ are tangent, normal and binormal unit vectors, respectively,
$k(\sigma)$, $\varkappa (\sigma)$ are the curvature and the torsion of the
string. Thus a curve can be equivalently characterized by $( k(\sigma)$,
$\varkappa (\sigma)$, $\vek{t} (0)$, $\vek{n} (0)$, $\vek{b} (0)$,
$\vek{x})$, where
$\vek{x}$ is the position vector of one of the two end - points of the string.
Solving the eq. \thetag{B1} one obtains another description of the curve:
$(\vek{t}(\sigma), \vek{n} (\sigma), \vek{b} (\sigma), \vek{x} ) $. If one
identifies $\vek{t} = \vekgr{\nu}_3$, $\vek{n} =\vekgr{\nu}_1$, $\vek{b}
=\vekgr{\nu}_2$, then a straight line given by $ k(\sigma) = 0$,
$\varkappa (\sigma)=0$ corresponds to a rotator. It is well known that in
this case the unit vectors $ \vek{n}$, $\vek{b}$ can be chosen arbitrarily
in a plane orthogonal to $\vek{t}$. This freedom corresponds just to the
axial symmetry group $U(1)$ of the rotator.

In quantum theory we expect that the state of a string (at least in a
simplified nonrelativistic version) is characterized by the wave functions
$$ \align \Phi_s & = \Phi_s[\vekgr{\nu}_1 (\sigma), \vekgr{\nu}_2 (\sigma), 
\vekgr{\nu}_3 (\sigma), \vek{x} ] \qquad \text{or} \\
\Phi_s & = \Phi_s[k(\sigma), \varkappa (\sigma), \vekgr{\nu}_1 (0), 
\vekgr{\nu}_2 (0), \vekgr{\nu}_3 (0), \vek{x} ] \endalign $$
In case of a frozen string $k(\sigma )$, $\varkappa (\sigma) $ do not depend
on time and ceased to be real degrees of freedom. Then the wave function
reduces to
$$ \Phi_s  = \Phi_s[\vekgr{\nu}_1 (0), \vekgr{\nu}_2 (0),
\vekgr{\nu}_3 (0), \vek{x} ] = \Phi (\xi_i , \vek{x})\ , $$
which is the wave function considered in this paper. We have seen that
$\Phi (\xi_i , \vek{x} ) $ can describe spin 1/2 particle, so we expect that
also in general situation corresponding to $\Phi_s $, half - integral spin
is allowed.
\vskip \aaa
\centerline{\bf Acknowledgements}
\vskip \bbb
I am very indebted to M. Fecko for his support. I want to thank M. Kras\v
nansk\'y, J. Gregu\v s, P. Weisenpacher, B. Majern\'ik and R. Turansk\'y for
their technical assistance and A. Turzov\'a for encouragements.

\vskip \aaa
\centerline{\bf References }
\vskip \bbb
\noindent
[1] M. Green, J. Schwarz and E. Witten, Superstring theory
    vol. 1, 2, Cambridge University Press (1987) \newline \noindent
[2] A. Candelas, M. Horowitz, A. Strominger and E. Witten:
    Nucl. Phys. {\bf B258}, 46 (1985). \newline \noindent
[3] L. Dixon, J. A. Harvey, C. Vafa and E. Witten, Nucl.
    Phys. {\bf B261}, 678 (1985), {\bf B274}, 285 (1986). \newline \noindent
[4] I. Antoniadis, J. Ellis, J. S. Hagelin and D. V. Nanopoulos:
    Phys. Lett. {\bf B194}, 231 (1987), {\bf B231}, 65 (1989).
    \newline \noindent
[5] A. B. Lahanas and D. V. Nanopoulos, Phys. Rep. {\bf 145},
    1 (1986). \newline \noindent
[6] M. Petr\'a\v s,  Czech. Jour. Phys. {\bf B39}, 1208 (1989).
    \newline \noindent
[7] M. Petr\'a\v s, Czech. Jour. Phys. {\bf 43}, 1147 (1993).
    \newline \noindent
[8] V. Balek, M. Melek and M. Petr\'a\v s, Czech. Jour. Phys. {\bf B37},
    1321 (1987). \newline \noindent
[9] M. Petr\'a\v s, Czech. Jour. Phys. {\bf 45}, 455 (1995). \newline \noindent
[10] M. Fecko, Czech. Jour. Phys. {\bf B35}, 378 (1985) \newline \indent
     M. Fecko, J. Math. Phys. {\bf 29}, 1079 (1988). \newline \noindent
[11] A. J. Hanson and T. Regge, Ann. Phys. {\bf 87}, 498 (1974).
     \newline \noindent
[12] F. Bopp and R. Haag, Z. Naturforschung {\bf A5}, 649 (1950).
     \newline \indent
     L. S. Schulman, Phys. Rev. {\bf 176}, 1558 (1968).
     \newline \noindent
[13] E. P. Wigner, in Theoretical Physics, IAEA Vienna (1963)
     \newline \noindent
[14] E. D. Commins and P. H. Bucksbaum, Weak Interactions
     of Leptons and Quarks, Cambridge University Press (1983)

\enddocument
\end